\documentclass[aps,prd,preprint,floatfix,nofootinbib]{revtex4}

\usepackage{graphicx,epsfig}

\def\dilog#1{\,{\rm Li_2}#1}                        

\def\eg{{\it e.g.}}
\newcommand{\GeV}{\unskip\,\mathrm{GeV}}

\newcommand{\TeV}{\unskip\,\mathrm{TeV}}

\newcommand{\gsim}
{\;\raisebox{-.3em}{$\stackrel{\displaystyle >}{\sim}$}\;}

\begin{document}

\begin{flushright}
OITS-763\\
hep-ph/0503116\\
\end{flushright}


\bibliographystyle{revtex}

\title{Signals for CP Violation in Split Supersymmetry}




\author{N.G. Deshpande}
\author{J. Jiang}
\affiliation{Institute of Theoretical Sciences, University of Oregon,
Eugene, OR 97403}


\date{\today}

\begin{abstract}
Split supersymmetry is characterized by relatively light chargino and
neutralino sector and very heavy sfermion sector.  We study the
consequence of CP violation in this scenario by evaluating two-loop
contributions to electric dipole moments of fermions from Higgs-photon
as well as $W$-$W$ diagrams.  These contributions add coherently and
produce electron and neutron electric dipole moments close to present
bounds.  We then explore Higgs production at a photon-photon collider,
and consider the feasibility of measuring CP violating $h\gamma\gamma$
coupling induced by chargino loops.  Methods of enhancing the
sensitivity are discussed.  For lower chargino masses and lower Higgs
boson masses, the effect of the CP violation can be observed with $90\%$
confidence level significance.
\end{abstract}

\maketitle


\section{Introduction}
Supersymmetry(SUSY) has been one of the most promising candidates for
the extension of the Standard Model(SM).  It provides an elegant
solution to the gauge hierarchy problem.  Recently a new scenario of
SUSY model was proposed, in which solution of the naturalness problem
is no longer required~\cite{Arkani-Hamed:2004fb}.  This scenario is dubbed
split SUSY because of the hierarchical mass difference between the
scalar and the fermionic superpartners.  The other two prominent
features of SUSY, gauge coupling unification and dark matter candidate
are retained in split SUSY.  By allowing the existence of fine-tuning,
the SUSY breaking scale can be relaxed to be much higher than $1$ TeV.
Subsequently, the heavier sfermion masses help to eliminate several
unpleasant aspects of SUSY, including excessive flavor and CP
violation, fast dimension-5 proton decay and the non-observation of the
lightest CP even Higgs boson.  Various aspects of phenomenology in the
split SUSY scenario have been explored in
Refs.~\cite{Arkani-Hamed:2004yi,ssphen1,Chang:2005ac}.

Split supersymmetry is characterized by relatively light ($100$
GeV-$1$ TeV) charginos and neutralinos and much heavier squarks and
sleptons.  In this note we further explore some of the consequences of
CP-violation in split SUSY.  We shall consider electric dipole moment
(EDM) of electron and quarks, and arrive at their value in split
SUSY versus in standard SUSY.  Similarly, we consider CP violating
coupling of the Higgs boson to photons, and examine the feasibility of
measuring this effect at a $\gamma\gamma$ collider.  We shall allow CP
violating phases in the SUSY potential to take values of $O(1)$, and
all suppressions of one-loop contribution is attributed to higher masses
of the supersymmetric particles.

Electric dipole moments of fermions arise at one loop in conventional
SUSY.  As squark and slepton masses exceed $5$ TeV and charginos and
neutralinos remains light, the one-loop contributions become
comparable to the two-loop contributions.  In split SUSY, the two-loop
contributions arise from a set of Higgs-photon diagram considered
before~\cite{Arkani-Hamed:2004yi}, as well as the $W$-$W$ diagram,
that we consider here~\footnote{As we were preparing to submit this
paper, we noticed a similar study by Chang, Chang and
Keung~\cite{Chang:2005ac} was submitted to the arXiv.}.  Allowing SUSY
parameters to have arbitrary complex values, we show that these two
contributions always add coherently.  The predicted values of the
electron EDM in particular set useful constraint on split SUSY mass
scale, and further improvement in measurements~\cite{Kawall:2003ga}
can provide strong constraints on the theory.  We similarly discuss
neutron EDM.

Another CP violation signal is through the study of the
$h\gamma\gamma$ coupling.  In the SM, the Higgs coupling to photons
arises predominantly through $W$-boson and top-quark loop, and is CP
conserving.  In supersymmetry, a CP violating coupling can arise
through chargino loop, provided the complex phases in the chargino
sector are non-zero~\cite{Choi:2002rc,Arkani-Hamed:2004yi}.  The CP
violating effect is similar to that in a two Higgs doublet model with
CP violating mixing of scalor and pseudo-scalar Higgs.  We extend the
work of Ref.~\cite{Choi:2002rc} to a more realistic level and examine
the sensitivity of measuring CP violation at a future $\gamma\gamma$
collider.

After the introductions, we discuss the EDM in Sec.~\ref{sec:edm} and
$h\gamma\gamma$ coupling in Sec.~\ref{sec:hgg}.  Our conclusions are
presented in Sec.~\ref{sec:con}.

\section{Electric dipole moments}
\label{sec:edm}
\begin{figure}[htb]
\centering
\includegraphics[width=12cm]{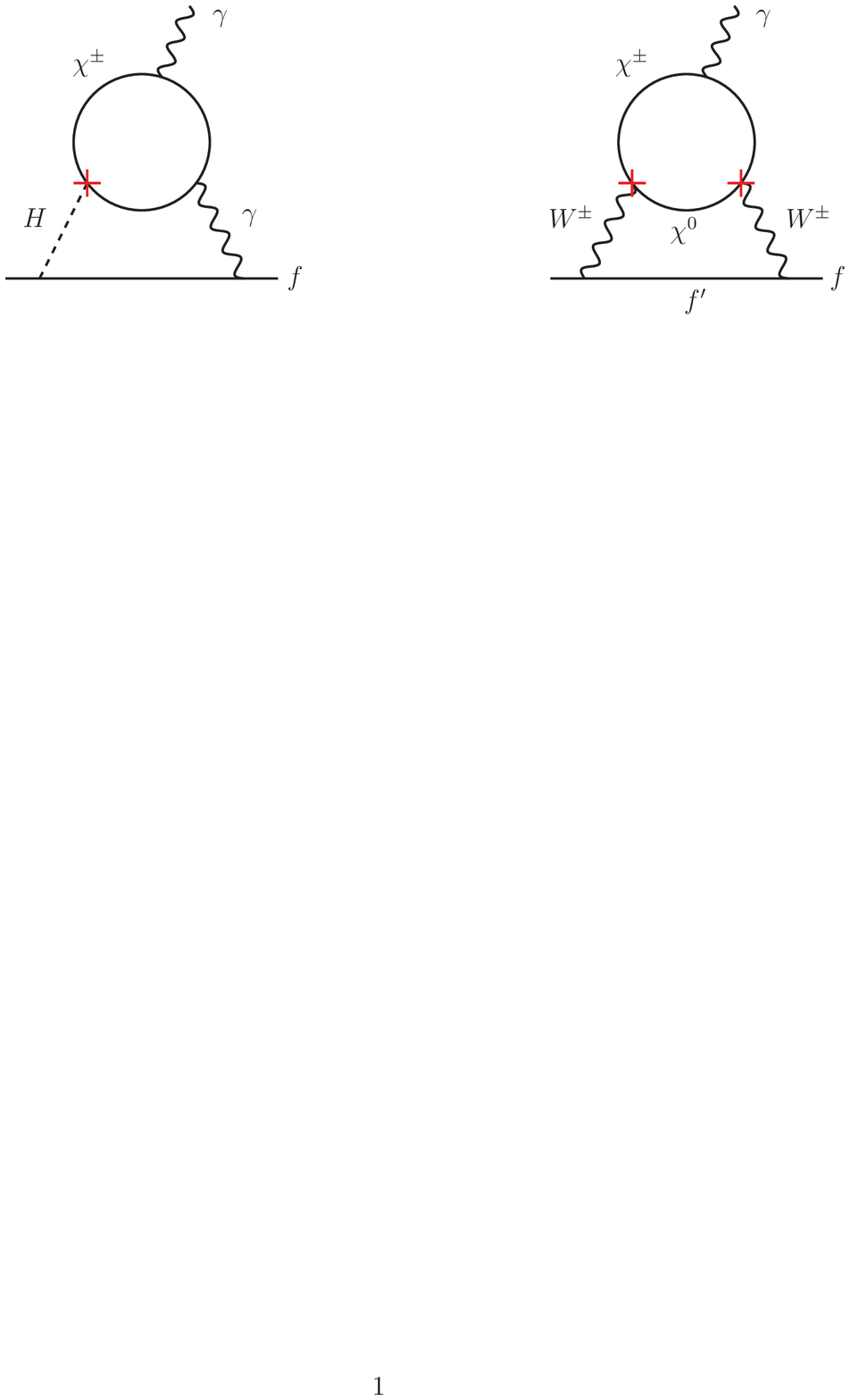}
\caption{Feynman diagrams of the fermion EDMs at two-loop level.  The (red) crosses
indicate CP violating couplings.}
\label{fig:feyn2loop}
\end{figure}

In split SUSY, as the sfermions get heavy, the one-loop contributions
to the fermion EDM get suppressed due to the large sfermion mass.  The
neutralinos, charginos and the lighter CP even Higgs boson remain
light.  The CP phases in the gaugino sector can induce EDM for
fermions at 2-loop level.  Study of the two-loop fermion EDMs in the
SM and SUSY can can be found in
Refs.~\cite{Barr:1990vd,Chang:1998uc,Chang:1999zw,Bowser-Chao:1997bb,Pilaftsis:2002fe,
West:1993tk,Kadoyoshi:1996bc}.  In split SUSY, the diagrams involving
charged Higgs bosons are suppressed due to the very large charged
Higgs boson masses.  The typical diagrams, shown in
Fig.~\ref{fig:feyn2loop}, include a set of diagrams that involve a
Higgs boson and a photon (left) and those that involve two $W$ bosons
(right).  The contributions from the Higgs diagrams have been studied in
Ref.~\cite{Arkani-Hamed:2004yi}.  We focus on the contributions from
the $W$-$W$ diagram.

To specify our notation, we start with the chargino and neutralino
mass matrices.  The chargino mass matrix is
\begin{eqnarray}
M_{\chi^+}=\left(
\begin{array}{cc}
M_2 & g v_2^* / \sqrt{2} \\
g v_1^* / \sqrt{2} & \mu
\end{array}
\right)~,
\end{eqnarray}
where $g$ is the weak coupling and $m_W$ is the $W$ boson mass.  In
general, the gaugino and higgsino mass parameter $M_2$ and $\mu$, and
the vacuum expectation values $v_1$ and $v_2$ are all complex. After
absorbing three of the complex phases through field redefinition,
there are only one independent phase $\phi_{\mu}$ left.  The chargino
mass matrix can be diagonalized by unitary matrices $U$ and $V$,
\begin{equation}
U^\dagger M_{\chi^+} V = {\rm diag}(m_{\chi_1^+},m_{\chi_2^+})~,
\end{equation}
with the chargino masses satisfying $m_{\chi_1^+} < m_{\chi_2^+}$.
The neutralino mass matrix depends on an additional gaugino mass parameter
$M_1$,
\begin{eqnarray}
M_{\chi^0} = \left(
\begin{array}{cccc}
M_1 & 0   & - g' v_1^* / 2 & g' v_2^* / 2 \\
0   & M_2 &   g v_1^* / 2  & - g v_2^* / 2 \\
-g' v_1^* / 2 & g v_1^* / 2 & 0 & - \mu \\
g' v_2^* / 2 & -g v_2^* / 2 & -\mu & 0 
\end{array}
\right)~,
\end{eqnarray}
and after redefinition of fields, two independent phases remain,
$\phi_1$ of $M_1$ and $\phi_{\mu}$.  From now on, we keep only the
complex phases $\phi_\mu$ and $\phi_1$ and set all other parameters to be
real.  The mass matrix above can be diagonalized by an unitary matrix N,
\begin{equation}
N^T M_{\chi^0} N = {\rm diag}(m_{\chi_1^0}, m_{\chi_2^0},
m_{\chi_3^0},m_{\chi_4^0})~,
\end{equation}
where the neutralino masses are in the order of $m_{\chi_1^0} <
m_{\chi_2^0} < m_{\chi_3^0} < m_{\chi_4^0}$.  In this notation, the
two loop contribution to fermion EDM from the Higgs-photon digram is,
\begin{eqnarray}
d_f^h &=& \frac{e\,\alpha\, Q_f\, m_f\, g^2}{32 \sqrt{2} \pi^3\, m_W \,m_h^2}
\left( 1-\frac{4\alpha}{\pi} {\rm ln}\frac{m_h}{m_f} \right) \nonumber
\\
&& {\rm Im} \sum_{j=1}^{2} (\cos\beta\, U_{i2}\, V_{i1} + \sin\beta\, U_{i1}\,
V_{i2} )\, m_{\chi_i^+} f\left( \frac{m_h^2}{m_{\chi_i^+}^2} \right)~,
\\
f(x) &=& \frac{2\sqrt{x}}{\sqrt{x-4}} \left[\ln x 
\ln\frac{\sqrt{x-4}+\sqrt{x}}{\sqrt{x-4}-\sqrt{x}} 
+ \dilog{\left( \frac{2 \sqrt{x}}{\sqrt{x} - \sqrt{x-4}} \right )} 
- \dilog{\left( \frac{2 \sqrt{x}}{\sqrt{x} + \sqrt{x-4}} \right )}
\right]~. \nonumber 
\end{eqnarray}
Here $Q_f$ and $m_f$ are the charge and mass of the fermion
respectively, $m_h$ is the mass of the lightest CP even Higgs boson
and $\tan\beta$ is the ratio of $v_2$ and $v_1$, $\tan\beta = v_2/v_1$.

To evaluate the $W$-$W$ diagram, it is necessary to write out the
Lagrangian involving $W$ boson, neutralinos and charginos
\begin{eqnarray}
{\cal L} = \frac{1}{\sqrt{2}} g \overline{\chi_i^0} \gamma^\mu 
	\left(
	L_{ij} \frac{1-\gamma_5}{2} + R_{ij} \frac{1+\gamma_5}{2}
	\right)
	\chi_j^+ W_\mu^- + H.C.~,
\end{eqnarray}
where the couplings $L_{ij}$ and $R_{ij}$ ($i = 1,2,3,4$ and $j =
1,2$) are 
\begin{eqnarray}
L_{ij} &=& \sqrt{2} N_{i2} V_{j1}^* + N_{i3} V_{j2}^*~, \nonumber \\
R_{ij} &=& \sqrt{2} N_{i2}^* U_{j1} - N_{i4}^* U_{i2}~,
\end{eqnarray}
and they have different complex phases.

For electron, up and down quarks, their masses and the masses of their
$SU(2)$ partners are much smaller than the $W$ boson mass.  It is safe
to neglect these small masses in the loop integrations.  Taking this
limit and following Ref.~\cite{Kadoyoshi:1996bc}, the EDM of a fermion
arising form the $W$-$W$ diagram in Fig.~\ref{fig:feyn2loop} can be
approximated by
\begin{eqnarray}
d_f^W \approx && \mp\ e \left( \frac{\alpha}{4 \pi \sin^2\theta_W} \right)^2
\sum_{i=1}^{4} \sum_{j=1}^{2} {\rm Im} (L_{ij}^* R_{ij}) 
\frac{m_{\chi_i^0}\, m_{\chi_j^+}\, m_f}{2 m_W^4}  \nonumber \\
&& \int_0^1 dx 
\frac{(1-x) m_W^2}{x\,m_{\chi_i^0}^2 + (1-x) m_{\chi_j^+}^2}
\ln \left[ \frac{x (1-x) m_W^2}{x\, m_{\chi_i^0}^2 +(1-x)
m_{\chi_j^+}^2} \right]~.
\end{eqnarray}
The minus/plus sign in front of the expression corresponds to
fermions with third component of their isospin being $1/2$ and $-1/2$
respectively. 

Before presenting the numerical results, a few comments are in order.
The CP violating $WW\gamma$ coupling can induce EDM for $W$ boson at
one-loop level.  Directly measuring the $W$ boson EDM is difficult.
It can be constrained by measuring the electron and neutron EDMs.
Unlike in the one-loop case, where the electron and neutron EDM values
are enhanced by large value of $\tan\beta$, the two-loop contributions
are suppressed as $\tan\beta$ increases.  We use $d^h$ to denote the
the Higgs-photon contribution and $d^W$ the $W$-$W$ contribution to
the electron EDM.  Both $d^h$ and $d^W$ decreases as $m_{\chi^+}$
increases, while $d^h$ is also reduced as $m_h$ gets larger.  With our
choice of independent complex phases, $d^h$ depends only on $\phi_\mu$
and $d^W$ depends on both $\phi_\mu$ and $\phi_1$.

\begin{figure}[htb]
\centering
\includegraphics[width=8cm]{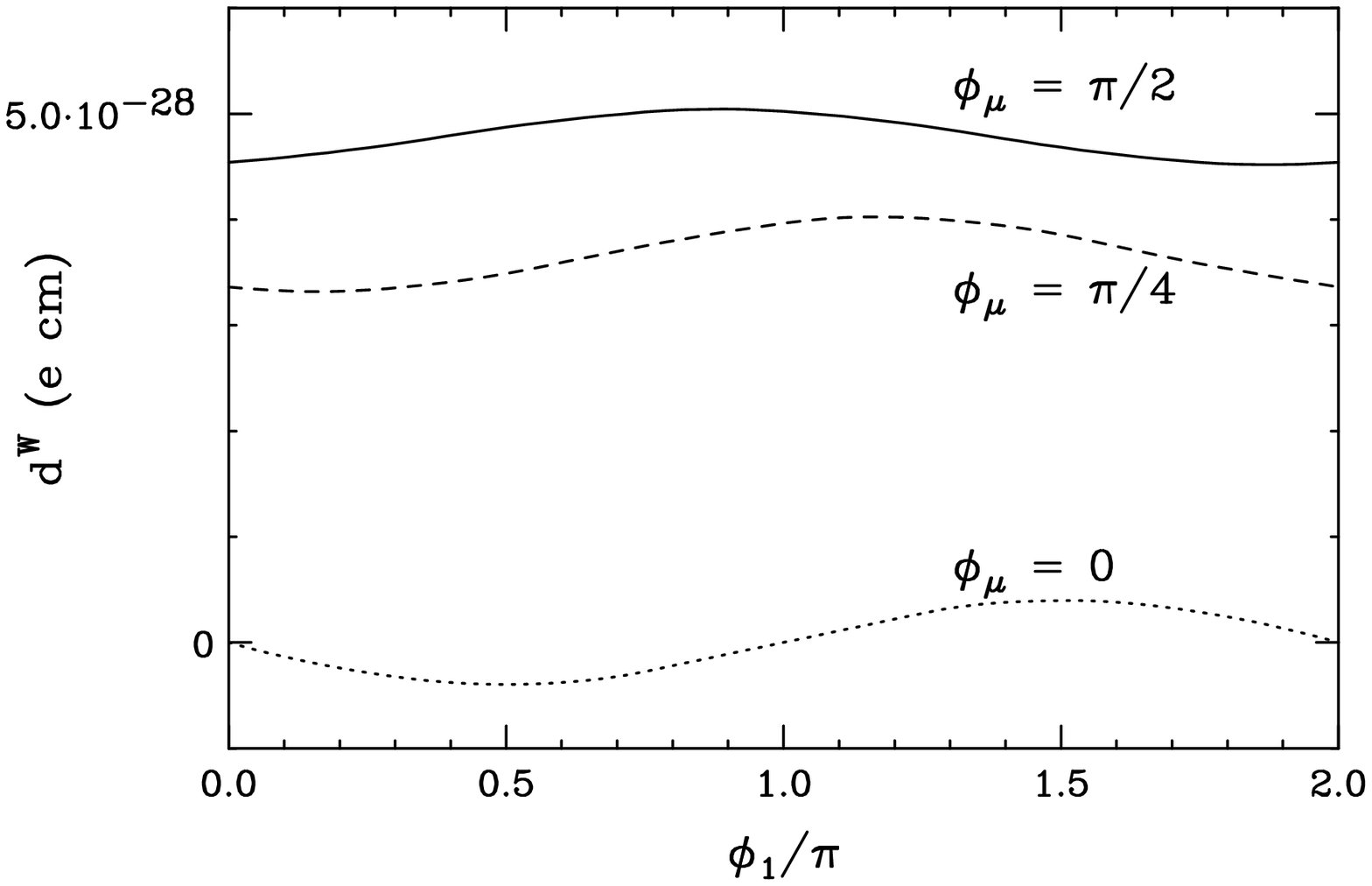}
\includegraphics[width=8cm]{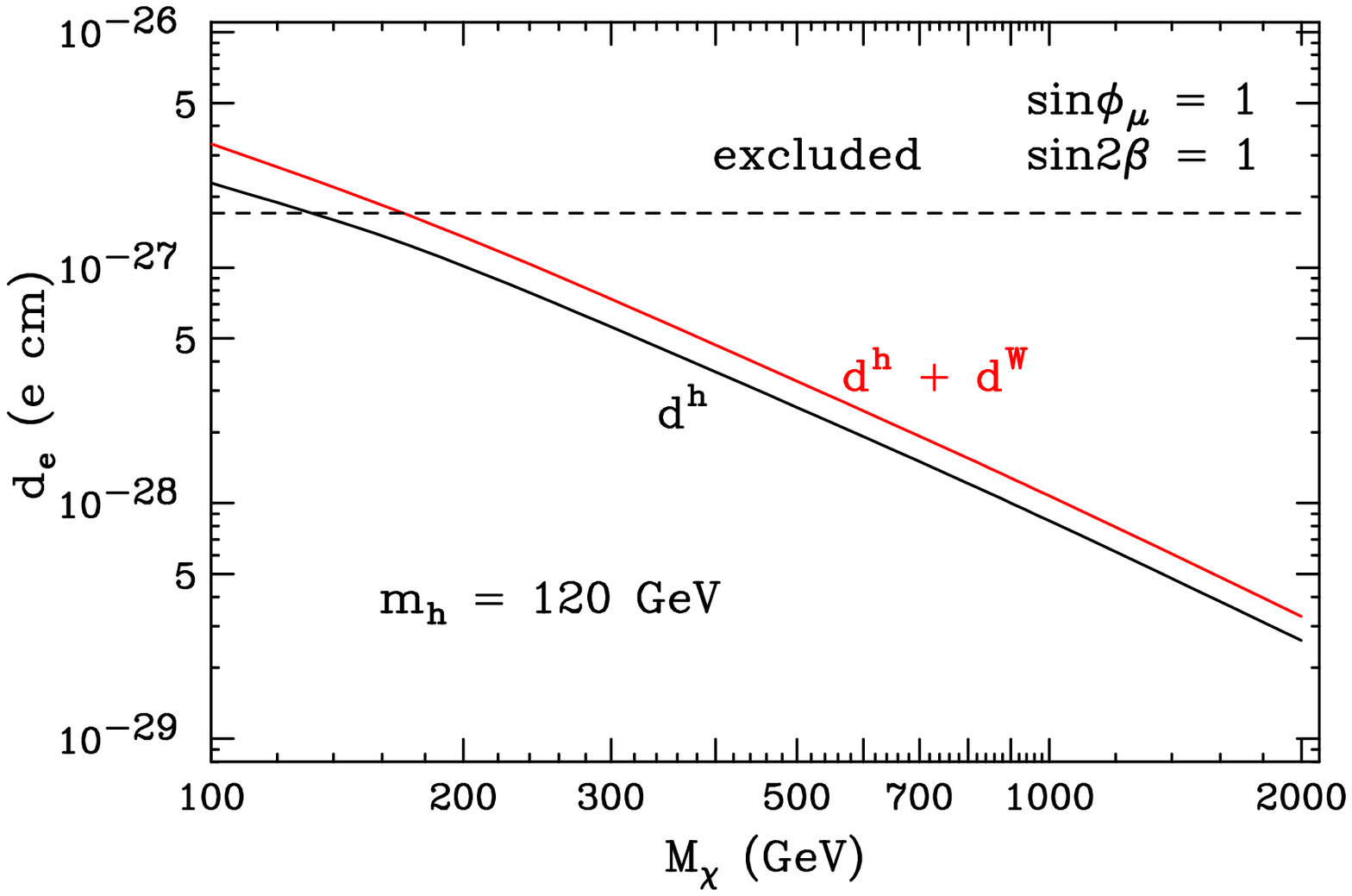}
\caption{Left: $W$-$W$ diagram contribution $d^W$ as a function of the
complex phase $\phi_1$ for $\phi_\mu = 0$ (dotted), $\pi/4$ (dashes)
and $\pi/2$ (solid).  Right: The dominant 2-loop contributions to the
electron EDM $d^h$ (black) and $d^h+d^W$ (red) as functions of
$m_{\chi_1^+}$, for $\phi_1 = 0$, $\phi_\mu = \pi/2$, $\tan\beta = 1$
$m_h = 120$ GeV and $m_{\chi_2^+}/m_{\chi_1^+} = 2$.}
\label{fig:edms}
\end{figure}

To show the dependence of $d^W$ on the complex phases $\phi_\mu$ and
$\phi_1$, we choose the following parameters for illustrative purpose
\begin{eqnarray}
&& |M_1| = 100 \GeV, \quad M_2 = 200 \GeV, \nonumber \\
&& |\mu| = 300 \GeV, \quad \tan\beta = 1.0~.
\label{eq:cond}
\end{eqnarray}
Although $d^W$ depends on both $\phi_1$ and $\phi_\mu$, the effect of
varying $\phi_\mu$ is more important.  We show $d^W$ as a function of
$\phi_1$ for $\phi_\mu = 0, \pi/4,$ and $\pi/2$ in the left panel of
Fig.~\ref{fig:edms}.  The variation of $d^W$ due to $\phi_1$ is an
order of magnitude smaller than the variation due to $\phi_\mu$.
Numerical evaluation also show that $d^h$ has the same sign as $d^W$
and is about twice in magnitude.  Thus, for large enough $\phi_\mu$,
independent of changes in $\phi_1$, $d^h$ and $d^W$ always add
constructively.

In the right panel of Fig.~\ref{fig:edms}, we show both the
contributions form the Higgs-photon diagram and the $W$-$W$ diagram.
Here, we use $\phi_1 = 0$, $\phi_\mu = \pi/2$, $\tan\beta = 1$, $m_h = 120
\GeV$ and the unification inspired mass relation $M_1 = 5/3 \tan^2
\theta_W M_2$ to reduce the number of variables.  As we vary $M_2$, we
change $\mu$ accordingly to maintain the chargino mass ratio
$m_{\chi_2^+}/m_{\chi_2^-} = 2$.  We see that the $W$-$W$ diagram
contribution is about $25\%$ to $50\%$ of that of the Higgs-photon
diagram for chargino mass range from $100$ GeV to $2$ TeV.  For larger
$m_h$, $d^h$ will be reduced, hence the relative importance of $d^W$
increases.  The dash line in the plot shows the current $95\%$
confidence level upper bound on the electron EDM, $|d_e| < 1.7 \times
10^{-27} {\rm e\,cm}$~\cite{Regan:2002ta}.  If the CP phases are
indeed of order $O(1)$, the electron EDM bound constraints the
chargino masses in split SUSY to be $m_{\chi_1^+} \gsim 150 \GeV$.
The next generation EDM experiments can improve the sensitivity by a
few order of magnitude~\cite{Kawall:2003ga}.  Again assuming order
$O(1)$ CP phases, these measurements will either observe the electron
EDM or put stronger constraints on chargino masses in split SUSY.

\begin{figure}[htb]
\centering
\includegraphics[width=8cm]{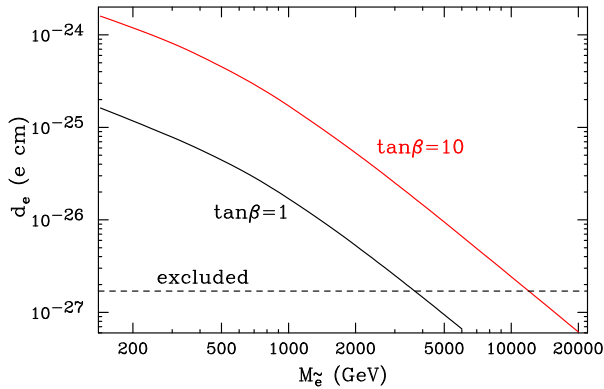}
\caption{One-loop electron EDM values as a function of the selectron mass.} 
\label{fig:sferm}
\end{figure}

If the sfermion masses are of the order of TeV, the one-loop diagrams
involving sfermions and gauginos or gluions will dominate the EDM
contribution.  In the MSSM, the predicted EDM values of the fermions
can be much larger then the current experimental bounds.  The fact
that we have not observed large EDMs can be explained by, small
complex phase, larger supersymmetric particle masses, cancellation at
work or a combination of the above~\cite{Ibrahim:1998je}.  If we
assume that the phases are of order $O(1)$ and no large cancellation is
present, the remaining explanation is to adopt heavy sfermion masses.
In Fig.~\ref{fig:sferm}, we plot the one-loop prediction of electron
EDM coming from the neutralino-selectron and chargino-sneutrino
diagrams as a function of the selectron mass, while the sneutrino mass
is set to be the same as the selectron mass.  We see that, for
$\tan\beta = 1$, a selectron mass of $m_{\tilde e}
\approx 5 \TeV$ is sufficient to suppress the electron EDM to be below
the experimental bound.  For $\tan\beta = 10$, the corresponding mass
is $m_{\tilde e} \approx 20 \TeV$.  If we compare Fig.~\ref{fig:sferm}
to the right panel of Fig.~\ref{fig:edms}, it is interesting to note
that, in the SUSY parameter space where sfermion masses are of a few
TeV and the gauginos are light, both the one-loop and the two-loop
contributions are equally important.

The same two-loop diagrams can also generate EDMs for up and down quarks,
when the one-loop contributions are suppressed by the large squark
masses.  The quark EDMs manifest through the EDM of neutron.  In the
conventional SUSY, when squarks are around $1 \TeV$, there are also
chromoelectric dipole moments and gluonic dipole moments.  As the
squarks become heavy, both one-loop and two-loop contributions from
these sources are suppressed.  Lacking the full knowledge of the
neutron wave function, we use the chiral quark model
approximation~\cite{Manohar:1983md} to estimate the neutron EDM from
the quark EDMs,
\begin{equation}
d_n = \frac{\eta_e}{3}(4 d^d - d^u)~,
\end{equation}
where $d^d$ and $d^u$ are the down quark and up quark EDMs and $\eta_e
\approx 1.53$ is the QCD correction factors.  We evaluate the two-loop
induced neutron EDM for the same set of parameters as in
Eq.(\ref{eq:cond}).  The estimated neutron EDM is $4.0 \times
10^{-26}{\rm e\,cm}$, which is close to the current experimental
$90\%$ confidence level upper bound of $6.3 \times 10^{-26} {\rm
e\,cm}$~\cite{Eidelman:2004wy}.  The predicted value will be smaller
for larger $\tan\beta$, $m_{\chi^+}$ and $m_h$, as in the case of
electron EDM.

\section{CP violation in $\gamma\gamma$ to $h$ production}
\label{sec:hgg}

The loop induced $h\gamma\gamma$ coupling in the SM is CP conserving.
However, if there exists mixing of the CP even and the CP odd Higgs
bosons, there would be CP violating $h \gamma\gamma$ coupling in two
Higgs doublet models. On the other hand, the chargino loop can induce
CP violating $h \gamma\gamma$ coupling due to the complex phases in
the chargino mass matrix.  In principal, this CP violation can
manifest in both the Higgs boson decay into two photons and production
of a Higgs boson in photon-photon collisions.  It is, in practice,
difficult to determine the helicities of the outgoing photons from the
Higgs decay.  The mixing of Higgs bosons of different CP state has
been discussed in Ref.~\cite{Grzadkowski:1992sa,Ellis:2004hw}.
Similar to these analysis, the chargino loop induced CP violation can
also be explored at a photon
collider~\cite{Choi:2002rc,Arkani-Hamed:2004yi}.  We study in more
detail the experimental observables and the backgrounds and estimated
the sensitivity in determining the CP violating coupling.

\begin{figure}[htb]
\centering
\includegraphics[width=8cm]{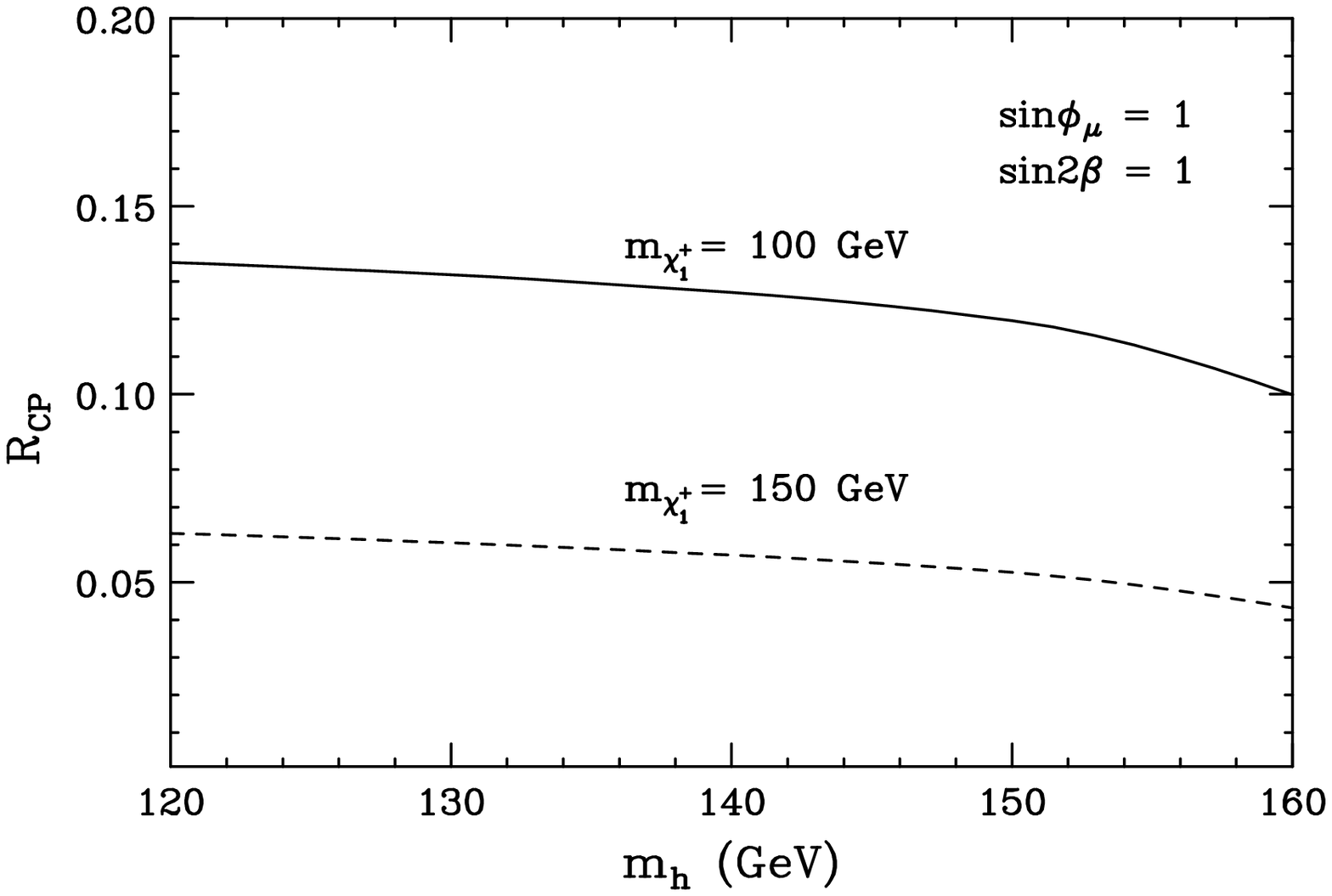}
\includegraphics[width=8cm]{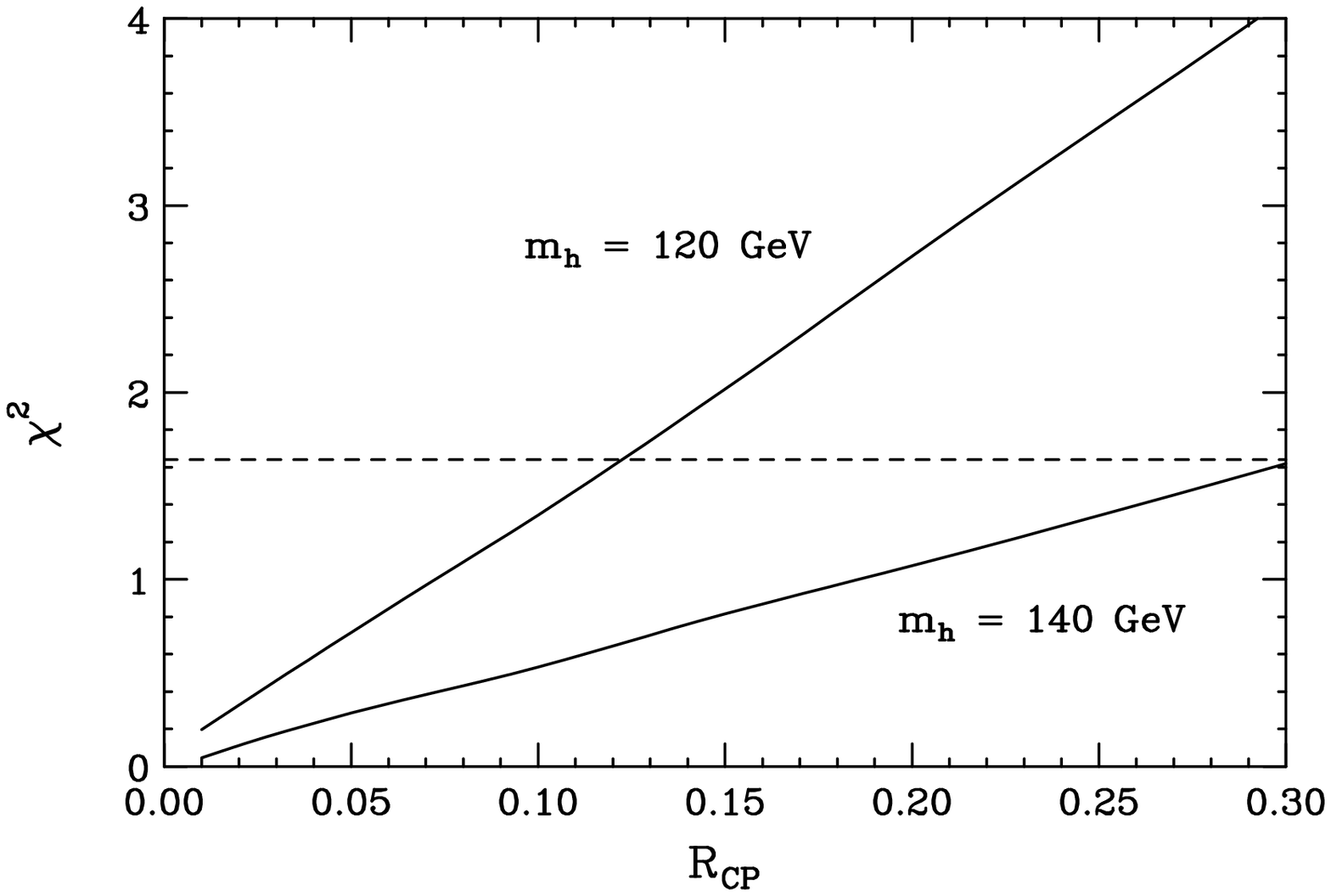}
\caption{Left: The ratio of $R_{CP}$ as a function of $m_h$.  The
solid curve is for $m_{\chi_1^+} = 100 \GeV$ and $m_{\chi_2^+} = 200
\GeV$ and the dashed curve for $m_{\chi_1^+} = 150 \GeV$ and
$m_{\chi_2^+} = 300 \GeV$.  Right: The statistical significance as a
function of $R_{CP}$ for $m_h = 120 \GeV$ and $m_h = 140 \GeV$.}
\label{fig:signif}
\end{figure}

The Higgs production rate in $\gamma\gamma$ collision is related to $h
\to \gamma \gamma$ decay width at a given $\gamma\gamma$ center-of-mass
energy $E_{\gamma\gamma}$ and the two colliding photon helicities,
$\lambda$ and $\lambda'$~\cite{Gunion:1989we}
\begin{equation}
\sigma(\gamma \gamma \to h \to X) = \frac{8 \pi \Gamma(h \to
\gamma\gamma) \Gamma(h \to X)}{(E_{\gamma\gamma}^2 -m_h^2)^2 +
\Gamma_h^2 m_h^2} (1 +\lambda \lambda')~,
\end{equation}
where $\Gamma(h \to X)$ is the partial width of Higgs boson decay to
$X$ and $\Gamma_h$ is the total decay width of the Higgs boson.  The $h \to
\gamma \gamma$ decay partial width is given by 
\begin{eqnarray}
&& \Gamma(h \to \gamma\gamma) = \frac{\alpha^2 g^2 m_h^3}{1024 \pi^3
m_W^2} (|e|^2 + |o|^2)~,  \\
&& e = \frac{4}{3} F_{1/2}\left( \frac{4 m_t^2}{m_h^2} \right) + F_1 \left(
\frac{4 m_W^2}{m_h^2} \right) \nonumber \\
&& \quad \quad \quad + \sqrt{2}\,{\rm Re}\sum_{i=1}^{2}(\cos\beta\,
U_{i2} V_{i1} + \sin\beta\, U_{i1} V_{i2}) \frac{m_W}{m_{\chi_i^+}} F_{1/2} \left(
\frac{4 m_{\chi_i^+}^2}{m_h^2} \right)~, \\
&& o =  \sqrt{2}\,{\rm Im} \sum_{i=1}^{2} (\cos\beta\,
U_{i2} V_{i1} + \sin\beta\, U_{i1} V_{i2}) \frac{m_W}{m_{\chi_i^+}} F_{1/2} \left(
\frac{4 m_{\chi_i^+}^2}{m_h^2} \right)~,
\end{eqnarray}
where the integration functions for spin-$1/2$ and spin-$1$ particles
in the loop are
\begin{eqnarray}
&&F_{1/2}(x) = - 2 x \left[ 1 + (1-x) \left( \arcsin
\frac{1}{\sqrt{x}} \right)^2 \right]~, \\
&&F_1(x) = 2 + 3 x \left[ 1 + (2-x) \left( \arcsin
\frac{1}{\sqrt{x}} \right)^2 \right]~.
\end{eqnarray}
Note because of the tininess of the bottom quark loop contribution, we
ignore it here.  The magnitude of the CP violation can be
characterized by the ratio $R_{CP} = |o/e|$.  We show $R_{CP}$ for
different chargino masses as a function of the Higgs boson mass in the
left panel of Fig.~\ref{fig:signif}.  $R_{CP}$ stays rather constant
for different values of $m_h$ until $m_h$ approach the threshold for
decay into two $W$ bosons.  Increasing $m_{\chi_1^+}$ to $150 \GeV$
will reduce $R_{CP}$ by about a factor of $2$.  For $m_{\chi_1^+} =
100 \GeV$, $\phi_\mu = \pi/2$, and $m_h = 120 \GeV$, $R_{CP}$ is about
$0.135$.

Three asymmetries can be constructed from $e$ and $o$
\begin{equation}
A_1 = \frac{-2 {\rm Im}(e o^*)}{|e|^2 + |o|^2}~, \quad
A_2 = \frac{-2 {\rm Re}(e o^*)}{|e|^2 + |o|^2}~, \quad
A_3 = \frac{|e|^2 - |o|^2}{|e|^2 + |o|^2}~.
\end{equation}
In the current case, both $e$ and $o$ are real, and $o$ is small
compared to $e$.  Therefore, $A_1$ is always $0$ and the deviation of
$A_3$ from $\pm 1$ is of order $(o/e)^2$.  The deviation of $A_2$ from
$0$ is of order $(o/e)$, thus rendering $A_2$ the most promising
observable.  The Higgs production rate can now be expanded in terms of
the asymmetries~\cite{Grzadkowski:1992sa,Choi:2002rc}
\begin{eqnarray}
d N = \frac{1}{2} d L_{\gamma\gamma}\, d \Gamma (|e|^2+|o|^2) [ (1 +
\langle \zeta_2 \zeta_2' \rangle ) + ( \langle \zeta_3 \zeta_1'
\rangle + \langle \zeta_1 \zeta_3' \rangle ) A_2 ]~,
\end{eqnarray}
where, $d L_{\gamma\gamma}$ is the luminosity of the back-scattered
photons, $d \Gamma$ is the phase space of the decay particles and he
$\zeta_i$ are the Stokes parameters, which indicate the degree of
linear and circular polarizations~\cite{Ginzburg:1981vm}.  In the
above expression we have dropped the $A_1$ term and we ``turn off''
the $A_3$ term by setting the azimuthal angle between the maximum
linear polarization direction of the two back-scattered photons
$\kappa$~\cite{Ginzburg:1981vm} to satisfy $\cos2\kappa = 0$.  The
quantity $A_2$ can be accessed by measuring the difference between the
production rates with $\sin2\kappa = -1$ and $1$.  To accentuate the
effect of $A_2$, it is preferable to make $( \langle \zeta_3 \zeta_1'
\rangle + \langle \zeta_1 \zeta_3' \rangle )$ as large as possible.
This is achieved by setting the ratio of the emitted photon energy to
the initial electron energy to be close to it is maximal
value~\cite{Grzadkowski:1992sa,Choi:2002rc}.  Thus the
electron-electron center of mass energy shall be slightly higher than
the Higgs threshold, \eg, $\sqrt{s_{ee}} = 150 \GeV$ for $m_h = 120
\GeV$ and $\sqrt{s_{ee}} = 175 \GeV$ for $m_h = 140 \GeV$.

As the Higgs boson in the mass range of $120-140$ GeV decays
significantly into $b\bar{b}$, we observe the Higgs boson production
signal in the $b\bar{b}$ final state.  Since it is only necessary to
tag one of the two $b$-jets, the tagging efficiency is $2 \epsilon_b
-\epsilon_b^2 \approx 98\%$, with $\epsilon_b = 85\%$ being the
tagging efficiency of one $b$-jet~\cite{Kuhlman:1996rc}.  There
exist a large $\gamma \gamma \to b\bar{b}$ and $c\bar{c}$ backgrounds.
Assuming the rate of mistagging a $c$-jet as a $b$-jet is $\epsilon_c
= 4.5\%$~\cite{Kuhlman:1996rc}, then the overall mistagging rate is $2
\epsilon_c -\epsilon_c^2 \approx 0.2\%$.  These two sources of
backgrounds can be significantly reduced by imposing the invariant
mass cut, $| m_{bb} - m_h | \le 10$ GeV and the angular cut on outgoing
$b$-jet direction relative to the beam line direction, $ 30^\circ <
\theta_{bz} < 150^\circ$.  With these cuts imposed, the background
cross sections are $\sigma_{bb} = 5.7$ fb and $\sigma_{cc} = 9.1$ fb.
As a comparison, for $m_h = 120 \GeV$ and $R_{CP} = 0.10$, the signal
cross section with $\sin2\kappa = 1$ is $\sigma_+ = 5.03$ fb and that
with $\sin2\kappa = -1$ is $\sigma_- = 4.66$ fb.  The total of $1\,{\rm
ab}^{-1}$ luminosity will be divided into $500\,{\rm fb}^{-1}$ for the
each of the $\sin2\kappa = -1$ and $1$ runs.  In our analysis, we use
$80\%$ initial electron polarization and $100\%$ polarization for
linearly polarized initial photons~\cite{Abe:2002wb}.

The statistical significance is presented by
\begin{equation}
\chi^2 = \frac{N_+ - N_-}{\sqrt{N_+ + N_- + 2 N_{BG}}}~,
\end{equation}
where $N_+$ and $N_-$ are the event number with $\sin2\kappa = 1$ and
$-1$ respectively and $N_{BG}$ is the sum of the $b\bar{b}$ and
$c\bar{c}$ background event numbers.  In the right panel of
Fig.~\ref{fig:signif} , we show the statistical significance as
a function of the ratio $R_{CP}$, where the dash line indicates the
significance corresponding to a $90\%$ confidence level measurement.
For $m_h = 120 \GeV$ and with a $1\,{\rm ab}^{-1}$ integrated
luminosity, $R_{CP} \approx 0.12$ can be observed with $90\%$
confidence level.  Thus for $m_{\chi_1^+} = 100 \GeV$ and $m_h = 120
\GeV$, the predicted $R_{CP} = 0.135$ can be observed with 
$90\%$ confidence level significance at a future $\gamma\gamma$
collider.  For $m_h = 140 \GeV$, the predicted $R_{CP} = 0.13$ is
harder to observe because of the reduced branching ratio of Higgs
boson decay to $b\bar{b}$.  Increasing luminosity will improve the
significance, as also including other channels of Higgs boson decay.

\section{Conclusion}
\label{sec:con}
We have explored the consequences of CP violation in split SUSY.
Assuming all CP phases are of $O(1)$, we find that fermion EDMs arise
from two loop diagrams in which gauginos are in the loops.  We have
shown that apart from Higgs-photon diagram already considered,
$W$-boson diagram is of comparable importance.  Furthermore, the two
diagrams always add constructively, and sum of their contributions are
close to the present experimental bounds.  In the case of the
electron, we already see that the present bound requires $m_{\chi_1^+}
\gsim 150 \GeV$, provided that phase $\phi_\mu \approx
\pi/2$.  An order of magnitude improvement in the electron EDM would
definitely constrain the chargino masses and would thus be competitive
with accelerator bounds.  We have also observed that unlike one loop
contribution, the two loop gaugino contribution is largest for small
$\tan\beta$.  We have compared with the one-loop sfermion
contribution, and we see that the contribution becomes small as
sfermion masses exceed several TeV, the precise value being a function
of $\tan\beta$.  We have also estimated the neutron EDM from two-loop
diagrams and find the predicted value close to the present bound, again
for choice of large complex phase.

Another consequence of the CP phase in the gaugino sector is the loop
induced CP violation in the $h\gamma\gamma$ coupling.  We have
considered studying this CP violating coupling at a future
$\gamma\gamma$ collider, by the measurement of the Higgs production cross
section for different initial photon polarizations.  We have optimized
the signal by arranging the initial electron and photon polarization
and minimized the background with kinematic cuts.  We conclude that
with a luminosity of $1\,{\rm ab}^{-1}$, for $m_{\chi_1^+} = 150
\GeV$, the lower $R_{CP} = 0.06$ might be difficult to observe.  A
$90\%$ confidence level observation of CP violation can be achieved
for $R_{CP} = 0.12$.  In the split SUSY the predicted value for
$m_{\chi_1^+} = 100 \GeV$ and $m_h = 120 \GeV$ is about $0.135$, thus
it is hopeful that this effect can be observed.  For higher Higgs
boson masses, it will be necessary to increase the luminosity and to
include other decay channels.

\begin{acknowledgments}
J.J. thanks Vernon Barger, Tao Han and Tianjun Li for helpful
discussions.  This research was supported by the U.S.~Department of
Energy, High Energy Physics Division, under Contract DE-FG02-96ER40969.
\end{acknowledgments}

\end{document}